\documentstyle[times,pramana,epsf,float]{ias}
\newcommand{\alphas}{\alpha_{\scriptscriptstyle S}}
\begin{document}
\mark{{WHEPP-6 QCD report}{WHEPP-6 QCD report}}

\title{WHEPP6 QCD Working Group Report}
\author{Coordinators: Sourendu Gupta${}^1$ and D.\ Indumathi${}^2$}
\author{Contributors: S.\ Banerjee${}^1$, R.\ Basu${}^2$, M.\ Dittmar${}^3$,
   R.\ V.\ Gavai${}^1$, F.\ Gelis${}^4$, D.\ Ghosh${}^1$, Sourendu Gupta${}^1$,
   D.\ Indumathi${}^2$, Asmita Mukherjee${}^5$}
\author{}
\address{1. Tata Institute of Fundamental Research, Homi Bhabha Road,
  Mumbai 400005, India.}
\address{2. The Institute of Mathematical Sciences, CIT Campus,
  Chennai 600113, India.}
\address{3. Eidgen\"ossische Technische Hochschule, ETH Z\"urich,
  CH-8093 Z\"urich, Switzerland.}
\address{4. Brookhaven National Laboratory, Nuclear Theory, Bldg 510A,
  Upton NY-11973, USA.}
\address{5. Saha Institute of Nuclear Physics, 1/AF Bidhan Nagar,
  Salt Lake City, Calcutta 700064, India.}
\keywords{QCD, polarised scattering, light front field theory, heavy ion physics,
   non-equilibrium field theory, parton distributions at LHC, fragmentation
   functions, event shapes.}
\pacs{
     11.10.Wx 
     12.38.-t 
     12.38.Mh 
     13.60.Hb 
     13.65.+i 
     13.87.Fh 
     }
\abstract{This is the report of the QCD working group at WHEPP 6. Discussions and
   work on heavy ion collisions, polarised scattering, and collider phenomenology
   are reported.}
\maketitle

\section{Introduction}
Sunanda Banerjee,
Rahul Basu,
Rajeev Bhalerao,
M.\ Dittmar,
Rajiv Gavai,
Fran\c{c}ois Gelis,
Dilip Ghosh,
Sourendu Gupta,
A.\ Harindranath,
D.\ Indumathi,
Rajen Kundu,
Kajori Majumdar,
Asmita Mukherjee,
Krishnendu Mukherjee,
R.\ Ratabole,
V.\ Ravindran,
H.\ S.\ Sharatchandra,
Ajit Mohan Srivastava,
W.\ L.\ van Neerven and
Raju Venugopalan
participated in the QCD working group.

The working group activity was structured around talks followed by discussions.
The emphasis in this working group was on these discussions, which were usually
moderated by the speaker.

In view of the forthcoming runs of RHIC, there is currently much interest in
the experimental programs of heavy-ion physics and spin-physics. There are groups
in the country which are actively interested in both kinds of physics. As a result,
a major group of talks focussed on heavy-ion physics and polarised hard scattering.
There were also discussions on QCD at colliders, light front QCD, non-equilibrium
field theory and the confinement mechanism in QCD. Some of these discussions have
already led to detailed work and publications, and some more work will come out
of this meeting.

A list of the discussion talks is given in Table \ref{tb.talks}.  The
talk by Harindranath is part of this proceedings, and
the content of the talks by Venugopalan and Srivastava have been included
in their contributions to this proceedings.  This report summarises the
rest of the talks and work by participants in this working group.

\goodbreak
\begin{table}[hbt]
\caption{List of discussion talks and speakers in the QCD working group.}
\begin{center}\hskip4pc\vbox{\columnwidth=24pc\begin{tabular}{ll}
\hline
Confinement in QCD & H.\ S.\ Sharatchandra\\
Light front QCD & A.\ Harindranath\\
Polarised gluon density & S.\ Gupta\\
LHC as a parton collider & M.\ Dittmar\\
Heavy-ion collisions & R.\ V.\ Gavai\\
Low-x gluons in nuclei & R.\ Venugopalan\\
Non-equilibrium field theory & F.\ Gelis\\
Disoriented chiral condensates & A.\ M.\ Srivastava\\
Thermalisation in heavy-ion collisions & R.\ Venugopalan\\
QCD in polarised scattering & W.\ L.\ van Neerven\\
Low-x fragementation & D.\ Indumathi\\
Light front QCD & A.\ Mukherjee \\
\hline\end{tabular}}\end{center}\label{tb.talks}\end{table}

\section{Determining polarised parton densities:
    D.\ Ghosh, Sourendu Gupta, D.\ Indumathi}

It is widely known that the polarised gluon density is not determined
well by current data on polarised deep inelastic scattering. This
is surprising. After all $\alphas$ is known to good accuracy after
LEP. Then knowledge of the variation with $Q^2$ of the structure function
$g_1(x,Q^2)$ should be enough to give us a good handle on the polarised
gluon density.

The argument is correct, but measurement errors are large in the
region where the slope of $g_1$ determines the polarised gluon density.
At present an NLO analysis can only determine the sign of the first moment
of the polarised gluon density (it is negative in the $\overline{\rm
MS}$ scheme).

Another surprise is that the world data on $g_1$ can be fitted equally
well by flavour SU(2) symmetric sea as by one in which this symmetry is
maximally violated. Again this is due to errors in $g_1$ at moderately
low values of $x$.

It turns out that improved polarised DIS measurements at $x<0.1$ would
be sufficient to remove a large part of the uncertainty in polarised
sea and gluon densities. Details have now been published in \cite{ggi}.

\section{Transverse spin and polarized DIS: Asmita Mukherjee}

The complexities of spin of a composite system in the equal time
formulation of relativistic quantum field theory is well known
\cite{alfaro}. The Pauli-Lubansky operators qualify for the spin operators
only in the rest frame of the particle. In an arbitrary frame the
situation is complex because of the complicated interaction dependence
arising from the dynamical boost generators in both longitudinal and
transverse spin operators and the difficulty of the separation of the
center of mass motion from the internal motion \cite{osb}.

Light front theory gives a unique opportunity to address the issue of
the relativistic spin operators in an arbitrary reference frame since
boost is kinematical in this formulation \cite{hari1}. The longitudinal
spin operator (light front helicity) is interaction independent and the
interaction dependence of the transverse spin operators arise solely
due to the transverse rotation operators \cite{ls78}.  We have derived
the transverse spin operators for QCD starting from the manifestly
symmetric, gauge invariant energy-momentum tensor in the light-front
gauge. These can be separated into three parts; the first part depends
on the coordinates explicitly, the other two parts ('intrinsic') come
from the fermionic and gluonic parts of the energy-momentum tensor
respectively. The fermionic part is directly related to the integral of
$g_T$ \cite{hari2}. In analogy with the helicity sum rule \cite{hari3},
we propose a transverse spin sum rule.

\section{Enhancement of intermediate mass range dimuons in A-A${}'$ collisions:
  R.\ V.\ Gavai}

$J/\psi$-suppression has been regarded as a clean probe of quark-gluon
plasma formation in heavy ion (A-A') collisions.  A lot of excitement
has been created recently by the announcement \cite{na50} of the CERN NA50
experiment, reporting anomalous suppression in Pb-Pb collisions.
$J/\psi$ is detected in NA50 (and in its predecessor NA38) as a peak in
dimuon spectrum at its known mass of 3.1 GeV.  One needs to understand
well the continuum $\mu^+  \mu^-$-spectrum in the mass range around the
$J/\psi$ mass in order to extract the $J/\psi$ cross section and then
check for suppression or otherwise.  The continuum spectrum is in itself
interesting as possible thermal effects may show up in it, giving
another window on the hot QGP or thermal matter.

NA50 reported an enhancement \cite{dimu} in the dimuon spectrum in the
intermediate mass range, defined as that between 1.6 and 2.5 GeV, for
S-U and Pb-Pb collisions.  Using PYTHIA 6.1 to get shapes of Drell-Yan
and open charm contribution (with different intrinsic $k_T$ for the two)
and $m_c = 1.5$ GeV, open charm cross section is obtained by fitting the
normalizations of these and $J/\psi$ and $\psi'$ terms to the data.
While this exercise gave open charm cross section for p-A which was in
agreement with other measurements, its extrapolation to the S-U and
Pb-Pb cases are lower than the data, with the enhancement in data
increasing as a function of $E_T$.

The working group discussed the possibilities that the excess may be due
to 1) the absorbed or broken $J/\psi$ in nuclear environment or 2)
thermal dileptons \cite{rapp} or 3) due to the extrapolation procedure in 
comparing the pA with AA'.  It is hoped that they will be taken up for detailed
studies in near future.

\section{Out-of-equilibrium field theory: F.\ Gelis}

A naive way to generalize thermal field theory to out-of-equilibrium
situations is to replace the Bose-Einstein and Fermi-Dirac distributions
by arbitrary functions describing the state of the system. The new
statistical functions are kept constant in time. This simplification seems
reasonable for systems that return very slowly to their equilibrium state,
in which one is interested by a very fast microscopic process. 

However, it was noticed in \cite{AltheS} that infinities known as ``pinch
singularities'' (and appear as products of $\delta$ functions with the
same argument) do not cancel if one uses this procedure, even for
arbitrary small departures from equilibrium, contradicting the heuristic
argument used to justify the procedure. 

In \cite{Althe}, it was shown that giving a decay width to the particles
would regularize this problem. \cite{Bedaq} showed that these
singularities do not appear if one takes into account the relaxation of
the system towards equilibrium. 

Recently, \cite{Niega} unified these two different solutions by showing
that the pinch singularities cancel if one let the statistical weights
evolve in time according to a Boltzmann equation (the relation with
\cite{Althe} is that the collision term of the Boltzmann equation is
responsible for the decay width of the particles). In this improved
procedure, the would-be singular terms are finite and of order $\tau_{\rm
micro}/\tau_{\rm relax}$ where $\tau_{\rm relax}$ is the relaxation time
of the system, and $\tau_{\rm micro}$ is the timescale of the microscopic
process one is studying, in agreement with the intuitive argument stated
before. 

\section{LHC as a parton collider: M.\ Dittmar}

A new approach to the luminosity, parton distribution functions and
cross section measurements at the LHC is proposed \cite{Dittmar:1997md}.
The proposal considers the LHC directly as parton--parton collider.
The combination of parton distribution functions and the proton--proton
luminosities has thus to be replaced with parton--parton luminosities,
which can be measured precisely from theoretically well understood
reactions \cite{Behner:1997es}.  Candidates for reactions which constrain
the quark and antiquark luminosities are the resonance production of
W and Z bosons with their leptonic decays. Gluon luminosities can be
constrained from events with high transverse momentum photons, W and
Z bosons which are dominated by the scattering of quarks and gluons
\cite{Chiappetta:1999yi,Kajari}.

Studies indicate that the above reactions can be measured with negligible
statistical errors and that experimental systematic uncertainties of
perhaps $\pm$ 1\% can probably be obtained up to rapidities of 2.5.
The analysis of the rapidity distributions for the above reactions
provides very accurate parton--parton luminosities for parton x ranges
between 0.0001 and 0.2 at $Q^{2} \approx 10^{4}$ GeV$^{2}$ and higher.

Following these optimistic experimental possibilites, and if future
theoretical calculations can achieve similar accuracies for other
processes relative to the control reactions, the LHC experiments can
give precision cross sections for various final states.

\section{Fragmentation functions at low-$x$: D.\ Indumathi}

Consider coherent time-like branching. If an external line emits a
gluon, this introduces a propagator factor that not only leads to
collinear enhancement, but also soft enhancement as the gluon energy
$\omega \to 0$. This factor leads to angular ordering of successive
branches and hence to jet formation. In short, when soft enhancements
are summed, they interfere destructively to reduce the phase space. The
jet production cross-section can be expressed in terms of the
fragmentation functions $D_i^h(x,Q^2)$ where $x = p\cdot q/Q^2$ is also
the momentum fraction of the fragmentating parton, $i$, carried by the
hadron $h$, in the parton model. The scaling violations of the
fragmentation functions can be expressed in terms of the DGLAP
evolution equations \cite{Dglap}, analogous to the case of parton
density distributions. We have
\begin{equation}
\frac{\partial}{\partial \log t} D_i (x,t) = \sum_i \int_x^1 \frac{{\rm
d}}{z} \frac{\alpha_s}{2\pi} P_{ji}(z,\alpha_s) D_j (x/z, t)~,
\end{equation}
where the splitting functions are perturbatively calculable as $P_{ji} =
P_{ji}^0 + (\alpha_s/(2\pi)) P_{ji}^1 + \cdots$. The effect of azimuthal
ordering can be incorporated by replacing
$t$ by $z^2 t$ in the argument for $D_j$ on the RHS of the above
equation \cite{Dok,Ellis}. This modified leading log approximation
(MLLA) leads to a gaussian form of the fragmentation function at
low-$x$ \cite{Dok,Bassetto}:
\begin{eqnarray}
xD(x,Q)= {N(Q)\over \sqrt{2\pi}\sigma(Q)} \exp\left[-\left[\log(x)
		-\log(x_0)\right]^2/[2\sigma^2(Q)]\,\right]~,
\end{eqnarray}
where $N(Q)$ is the total multiplicity, $\sigma$ is the width of the
gaussian and ${x_0}$ is the position of the peak of the gaussian. The $Q^2$
dependence of $N$, $\sigma$ and $x_0$ are computable for total inclusive
hadrons as an expansion in terms of the scale parameter, $Y =
\log(Q/\Lambda)$, $\Lambda = 200$ MeV \cite{Dok}.

Comparison with data at $e^+\, e^-$ colliders \cite{inclusive} as well
as at the HERA $e \, p$ collider \cite{hera} shows good agreement with
predictions for the scale dependence of the multiplicity as well as
the peak position of the Gaussian. Semi-inclusive data on octet baryon
and meson production at $e^+ \, e^-$ colliders are also well-described
\cite{octet} in this MLLA approximation.  The main motivation was to
discuss the approximations involved and validity of application of the
theory to collider data.

\section{Event shapes and power corrections in QCD: R.\ Basu and S.\ Banerjee}

Power Corrections to the leading twist  results of perturbative QCD are
being studied, specifically in the context of event shape variables
like the thrust. The work of  the Milan group gives a systematic
methodology of analytically continuing the strong coupling  constant to
low values of $Q^2$ using a spectral representation for $\alphas(Q^2)$
\cite{milan}. They use it to calculate the leading power corrections to
various measurables like hadronic decay widths, structure  functions,
event shapes in jet cross sections etc. They fit it to data to estimate
one of the  free parameters in their analysis (called the Milan
factor). However calculation of the power corrections, particularly in
event shapes like the thrust by including the  effect of quark masses
was carried out in \cite{rbasu}.

Rahul Basu and Sunanda Banerjee used recent measurements  of the thrust,
coupled  with the earlier data and include quark mass corrections from the
above paper (particularly for the c and b quarks), to reanalyse and get
a fresh estimate of the Milan factor It appears from preliminary analysis
that there is a substantial shift in the value of the Milan factor.

Some preliminary details can be  presented.  The average value of  1-T
for the massless  and massive case differ by  about 18\% at the lowest
value of $E_{cm}$. The defect  between this  and the data is presumed
to be made up by power correction as given by the Milan  group. A fit
to the power corrections has been done both  for the Milan factor and
for $\alphas$.

The following 2 fits were done--- in the first $\alphas$
and $\alpha_0$ were floated for massless and massive formula. For massless
formula the following results were obtained with Milan factor of 1.795---
\begin{center}\begin{tabular}{ccc}
 $\chi^2\qquad$ & 100.532  & $\qquad$for 35 points \\
 $\alphas\qquad$ & 0.14757 & $\qquad0.16916\times10^{-2}$ \\
 $\alpha_0\qquad$ & 0.72956 & $\qquad0.56922\times10^{-2}$ \\
\end{tabular}\end{center}
For the massive formula the results were---
\begin{center}\begin{tabular}{ccc}
 $\chi^2\qquad$ & 101.138  & $\qquad$for 35 points \\
 $\alphas\qquad$ & 0.15035 & $\qquad0.16034\times10^{-2}$ \\
 $\alpha_0\qquad$ & 0.72057 & $\qquad0.63981\times10^{-2}$ \\
\end{tabular}\end{center}
In the second case $\alphas$, $\alpha_0$ were not floated in the massive
formula but the Milan factor was. Then the result is---
\begin{center}\begin{tabular}{ccc}
 $\chi^2\qquad$ & 107.285  & $\qquad$for 35 points \\
 ${\rm Milan}\qquad$ & 1.7116 & $\qquad0.21051\times10^{-1}$ \\
\end{tabular}\end{center}
A more detailed analysis is being carried out and will be reported elsewhere.


\begin{thebibliography}{99}
\bibitem{ggi}
   D.\ K.\ Ghosh, S.\ Gupta and D.\ Indumathi, hep-ph/0001287.
\bibitem{alfaro}
   V.\ de Alfaro, S.\ Fubini, G.\ Furlan, and G.\ Rossetti, {\sl Currents
   in Hadron Physics} (North-Holland, Amsterdam, 1973);
   C.\ Bourrley, E.\ Leader and J.\ Soffer, {\sl Phys.\ Rep.\/} 59 (1980) 95.
\bibitem{osb}
    See for example H.\ Osborn, {\sl Phys.\ Rev.\/} 176 (1968) 1514.
\bibitem{ls78} 
    K.\ Bardakci and M.\ B.\ Halpern, {\sl Phys.\ Rev.\/} 176 (1968) 1686;
    D.\ E.\ Soper, {\sl Field Theories in the Infinite Momentum Frame}, 
    Ph.\ D.\ Thesis, Stanford University, 1971, SLAC-137;
    H.\ Osborn, {\sl Nucl.\ Phys.\/} B 80 (1974) 90; 
    H.\ Leutwyler and J.\ Stern, {\sl Ann.\ Phys.\/} 112 (1978) 94. 
\bibitem{feza}
    F.\ Gursey in C.\ DiWitt and R.\ Omnes (eds.), {\sl High Energy
    Physics}, Gordon and Breach Science Publishers, 1965.
\bibitem{hari1}
   A.\ Harindranath, ``An Introduction to Light Front Dynamics for Pedestrians''
   and the references therein.
\bibitem{hari2}
   A.\ Harindranath, A.\ Mukherjee and R.\ Ratabole, hep-ph/9908424, to appear
   in {\sl Phys.\ Lett.\/} B.
\bibitem{hari3}
   A.\ Harindranath and R.\ Kundu, {\sl Phys.\ Rev.\/} D 59 (1999) 116013.
\bibitem{na50}
   M.\ C.\ Abreu {\sl et al.\/}, (NA50 Collaboration) CERN-EP-2000-013,
   {\sl Phys.\ Lett.\/} B, in press; {\sl Phys.\ Lett.\/} B 450 (1999) 456;
   {\sl ibid.\/}, B 410 (1997) 337.
\bibitem{dimu}
   M.\ C.\ Abreu {\sl et al.\/}, (NA50 Collaboration) CERN-EP-2000-012,
   {\sl Euro.\ Phys.\ J.\/} C, in press; {\sl J.\ Phys.\/} G 25 (1999) 235.
\bibitem{rapp}
   R.\ Rapp and E.\ Shuryak, {\sl Phys.\ Lett.\/} B 473 (2000) 19.
\bibitem{AltheS}
   T.\ Altherr and D.\ Seibert, {\sl Phys.\ Lett.\/} B 333 ({1994}) 149.
\bibitem{Althe}
   T.\ Altherr, {\sl Phys.\ Lett.\/} B 341 ({1995}) 325.
\bibitem{Bedaq}
   P.\ F.\ Bedaque, {\sl Phys.\ Lett.\/} B 344 ({1995}) 23.
\bibitem{Niega}
   A. Niegawa, {\sl Prog.\ Theor.\ Phys.\/} 102 ({1999}) 1.
\bibitem{Dittmar:1997md}
   M.\ Dittmar, F.\ Pauss and D.\ Zurcher, {\sl Phys.\ Rev.\/} D 56 (1997) 7284.
\bibitem{Behner:1997es}
   F.~Behner, {\sl Talk given at International Europhysics Conference on High-Energy
   Physics (HEP 97), Jerusalem, Israel, 19-26 Aug 1997\/}.
\bibitem{Chiappetta:1999yi}
   P.\ Chiappetta, G.\ J.\ Gounaris, J.\ Layssac and F.\ M.\ Renard,
   {\sl Phys.\ Rev.\/}  D 59 (1999) 014016.
\bibitem{Kajari}
   M.\ Dittmar and K.\ Mazumdar, {\sl The possibility to measure the charm, beauty
   and strange quark luminosity at the LHC\/}, CMS note in preparation.
\bibitem{Dglap}
   V.\ N.\ Gribov and L.\ N.\ Lipatov, {\sl Sov.\ J.\ Nucl.\ Phys.\/} 15 (1972) 438;
   Yu.\ L.\ Dokshitzer, {\sl Sov.\ Phys.\ JETP\/} 46 (1977) 641;
   G.\ Altarelli and G.\ Parisi, {\sl Nucl.\ Phys.\/} B 126 (1977) 298. 
\bibitem{Dok}
   Yu.\ L.\ Dokshitzer, V.\ A.\ Khoze, A.\ H.\ Mueller and S.\ I.\ Troyan, 
      {\sl Rev.\ Mod.\ Phys.\/} 60 (1988) 373;
   Yu.\ L.\ Dokshitzer, V.\ A.\ Khoze, A.\ H.\ Mueller and S.\ I.\ Troyan, 
      {\sl Basics Of Perturbative QCD\/} (Editions Frontieres) 1991.
\bibitem{Ellis}
   R.\ K.\ Ellis, W.\ J.\ Stirling and B.\ R.\ Webber, {\sl QCD and Collider Physics\/} 1990.
\bibitem{Bassetto}
   A.\ Bassetto, M.\ Ciafaloni and G.\ Marchesini, {\sl Phys.\ Rep.\/} 100 (1983) 201;
\bibitem{inclusive}
   W.\ Braunschweig {\sl et al.\/}, (TASSO Collaboration) {\sl Z.\ Phys.\/} C 47 (1990) 198;
   M.\ Z.\ Akrawy {\sl et al.\/}, (OPAL Collaboration) {\sl Phys.\ Lett.\/} B 247 (1990) 617.
\bibitem{hera}
   N.\ H.\ Brook, (For the H1 and ZEUS Collaborations) Talk presented at the second
   Latin American Symposium on high energy physics, SILAFAE98, San Juan, hep-ex/9805031.
\bibitem{octet}
   D.\ Indumathi, H.\ S.\ Mani, Anubha Rastogi, {\sl Phys.\ Rev.\/} D 58 (1998) 094014.
\bibitem{milan}
   Yu.\ L.\ Dokshitzer and B.\ R.\ Webber, {\sl Phys.\ Lett.\/} B 352 (1995) 451;
   Yu.\ L.\ Dokshitzer, G.\ Marchesini and B.\ R.\ Webber, {\sl Nucl.  Phys.\/} B 469 (1996) 93;
   Yu.\ L.\ Dokshitzer, A.\ Lucenti, G.\ Marchesini, G.\ P.\ Salam, {\sl Nucl.\ Phys.\/} B 511 (1998) 396.
\bibitem{rbasu}
   Rahul Basu, {\sl Phys.\ Rev.\/} D 29 (1984) 2642.
\end{thebibliography}
\end{document}